\documentclass[[preprint]{elsarticle}
\usepackage{amsmath}
\usepackage{amssymb}
\usepackage[tight]{units}
\usepackage{textcomp}
\usepackage[percent]{overpic}
\usepackage{color,graphicx}
\usepackage{natbib}
\usepackage{subfig}

\newcommand\geant{\textsc{Geant4}\/}
\newcommand{\dd}{\mathrm{d}}

\journal{Astroparticle Physics}

\begin{document}

\begin{frontmatter}

\title{Surface roughness interpretation of 730 kg days CRESST-II results}
\author[]{M.~Ku\'zniak\corref{cor1}}
\cortext[cor1]{Corresponding author.}
\ead{kuzniak@owl.phy.queensu.ca}
\author{M.~G.~Boulay}
\author{T.~Pollmann}
\address{Department of Physics, Queen's University, Kingston, Ontario, K7L 3N6, Canada}
\begin{abstract}
The analysis presented in the recent publication of the CRESST-II results~\cite{Angloher2009_run32}
finds a statistically significant excess of registered events over known background contributions
in the acceptance region and attributes the excess to a possible Dark Matter signal, caused by scattering of rather light WIMPs.
We propose a mechanism which explains the excess events
with ion sputtering caused by $^{206}$Pb recoils and $\alpha$-particles from $^{210}$Po decay, combined with realistic surface roughness effects.

\end{abstract}

\end{frontmatter}

\section{Introduction}
The experimental search for Dark Matter has recently entered a new exciting phase, reaching sufficient sensitivity to directly challenge some of
the proposed theoretical extensions of the Standard Model. One of the Dark Matter candidates, which has attracted particular attention is
the so-called Weakly Interacting Massive Particle (WIMP). More than a dozen experiments worldwide are currently aiming at direct detection of this 
class of particles. 
Current best limits on WIMP scattering cross-section have been reported by XENON100~\cite{XENON1002011_DMresults} and CDMS-II~\cite{CDMS}.
No unambiguous evidence for the existence of WIMPs has been presented so far, however several experiments (DAMA~\cite{DAMA}, for detailed interpretation see also~\cite{Savage2009_DAMAcompatibility}, more recently 
CoGeNT~\cite{Aalseth2011_CoGeNTModulation} and CRESST~\cite{Angloher2009_run32}) reported
statistically significant excesses of the number of detected events over expected contribution from all known backgrounds, which, under certain assumptions 
could be explained with the WIMP hypothesis.

Thorough understanding of all backgrounds is clearly necessary for the correct interpretation of results -- especially when the experiment is not background 
free (and all of the currently running Dark Matter experiments are background limited).
Typical sources of backgrounds include:
1) $e/\gamma$ events 2) neutrons 3) $\alpha$-particles and 4) nuclear recoils, with the last two often caused by $\alpha$-activity intrinsic to the detector
or coming from radon gas and its decay daughters. $^{222}$Rn with a half-life of 3.8~days is a particularly problematic background, since it can leave its decay 
products on surfaces, implant them just under the surface or even diffuse into materials before it decays. The most problematic radionuclide from the $^{222}$Rn 
chain is $^{210}$Pb with a half-life of 22.3 years and its decay daughter $^{210}$Po, which itself is an $\alpha$-emitter with a half-life of 138 days.  

In this article, we focus on the recent result from the CRESST-II Dark Matter Search and show that one of the four main backgrounds in the experiment, $^{206}$Pb 
recoils from $^{210}$Po decay, was possibly underestimated. Qualitatively, taking the $^{210}$Po decay into account with realistic models for surface roughness
(that have been motivated by our work in support of the DEAP~\cite{DEAPpaper,DEAP-1, TinaPhD} program), explains the excess events.

\section{CRESST-II detector}
The CRESST-II project (Cryogenic Rare Event Search with Superconducting Thermometers)~\cite{Angloher2009_run32, Angloher2009_run30, Angloher2005_cresstIIproof}, 
located at LNGS in Italy, uses cryogenic detector modules
which consist of scintillating CaWO$_4$ crystals operated as bolometers (sensitive to phonons) and a nearby but separate cryogenic light detector 
(a silicon wafer 
with a tungsten thermometer). Simultaneous detection in both the ``phonon channel'' and the ``light channel'' allows for efficient discrimination
between electronic, $\alpha$-particle and nuclear recoils, because of substantial differences in light yield. 

In addition to this, the crystal together with the light detector are completely surrounded by scintillating and reflecting foil. This is critical for 
discrimination of $^{210}$Po decays, which occur on the inner surface of the reflective housing or on the crystal surface. When the recoiling $^{206}$Pb enters 
the crystal, potentially generating a WIMP-like background event, the escaping \unit[5.3]{MeV} $\alpha$-particle reaches the foil and generates enough light 
to veto the event.

As discussed in the CRESST paper, the only non-scintillating surfaces inside the detector modules are the small silver-coated bronze clamps holding 
the target crystals, which have been identified as the main source of $\alpha$'s and Pb nuclear recoils. This detail is critical for the rest of this discussion.

\section{Analysis from the CRESST-II paper}
The CRESST-II paper extensively covers details of the maximum likelihood analysis applied to the final energy spectrum from the entire dataset. 
The final maximum likelihood fit is based on a detailed model including parametrized contributions from all backgrounds (as listed earlier) 
and a possible WIMP signal.
For each background, a reference region (i.e. range of energies and light yields) is defined. A data-driven model or approximated analytical function is developed
to describe the events in the reference region and eventually used to extrapolate the number of background events ``leaking'' into the 
acceptance region.
   
For the $e/\gamma$-band a Gaussian model is employed, which parametrizes the observed energy-dependent light distribution. Because of the high statistics available
for this class of events the extrapolation to the acceptance region is fairly accurate.

Simple linear functions with parameters constrained by the data are used to extrapolate the $\alpha$-particle spectra, separately for each detector module. The 
spectra are approximately flat over both the reference and acceptance regions and the extrapolation is fairly independent on the choice of the reference region.

A more sophisticated model, which includes hit multiplicities in more than one detector, is employed to account for neutrons of cosmogenic origin and 
for neutrons coming from natural radioactivity in the surroundings of the detector.

Finally, the $^{206}$Pb recoil background spectrum is modelled in a manner similar to the case of $\alpha$-particles. A clear peak from the full $^{206}$Pb deposited 
energy of \unit[103]{keV} is present in the spectrum, with an exponential tail extending towards lower energies. In the specified reference region
between 40 and \unit[90]{keV}, the tail is modelled by an exponential on top of a constant contribution:
\begin{equation} \label{eq:leadspectrum}
   \frac{\dd N_\text{Pb}}{\dd E}(E) = A_\text{Pb}^d \cdot
\left[C_\text{Pb}
+\exp\left(\frac{E-\unit[90]{keV}}{E^\text{Pb}_\text{decay}}
\right) \right],
\end{equation}
where $A_\text{Pb}^d$ may be module-dependent (to describe the
different recoil background rates in the individual detector
modules). The same spectral decay length
$E^\text{Pb}_\text{decay}$ and constant background term
$C_\text{Pb}$ are used for all modules. As noted by the authors:
``the latter quantities are
characteristic of the implantation profile of $\alpha$-emitters in
the clamps and can thus be assumed to be universal if the
underlying implantation mechanism is the same for all clamps''.

The authors used the SRIM
package \cite{SRIM_software} to simulate the  energy spectra
expected for different depth distributions of the 
$^{210}$Po parent nucleus. The simulated spectrum which best fitted the 
experimental spectrum in the reference region corresponded to the 
depth profile of the $^{210}$Po parent nuclei peaking at the clamp surface with 
a \unit[3]{nm} exponential decay length. It was concluded that none of the 
simulated spectra rose significantly towards lower energies within the range 
of the acceptance region, which then justified a simple exponential extrapolation 
of the Pb background.

In the next section we will show that the model with a simple exponential tail does not
correctly describe the $^{210}$Po induced spectrum and leads to a significant underestimation of the 
background contribution to the acceptance region.

\section{Lead recoil background}
\subsection{CRESST-II background model}
\label{sec:leadrecoils_qualitative}
We start by repeating the key SRIM simulation from Fig.~3 in the CRESST paper, which is the main justification
for the extrapolation. Decays of $^{210}$Po, i.e. a \unit[5.3]{MeV} $\alpha$ and a \unit[103]{keV} $^{206}$Pb recoil nucleus, emitted back-to-back, 
are generated in the clamp material according to an exponential depth profile with \unit[3]{nm} decay length 
peaking at the surface. 

Silver is the relevant base material for the simulation of $^{206}$Pb recoils coming from the clamps, as the thickness of the coating 
(several micrometers)
is significantly larger than the range of recoils in the material. In addition to this, simulations were also performed for bronze.
For bronze we simply used a ``typical'' composition provided in the database supplied with SRIM: 89\%~Cu, 9\%~Zn and 2\%~Pb, 
with \unit[8.82]{g/cm$^3$} density. For silver, natural abundances and \unit[10.5]{g/cm$^3$} density were used.

Since no evidence for material dependence was seen in any of the cases simulated for this work and the energy spectra, as well as the fit results, 
were within errors consistent for both silver and bronze, for the sake of clarity we only present detailed results for silver.

\begin{figure}
 \centering
 \includegraphics[width=\linewidth]{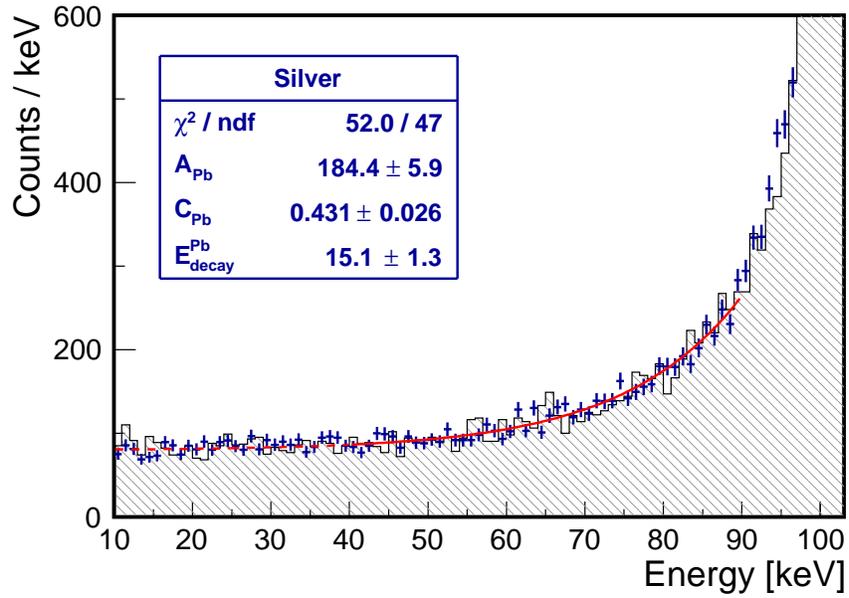}
 \caption{Energy spectrum of $^{206}$Pb recoils calculated with the
SRIM package for a depth distribution peaking at
the surface with a \unit[3]{nm} exponential decay length:
(shaded black histogram) plot digitized from Fig.~10 in the CRESST paper, (dark blue points) this work, with silver as the clamp material.
The red curve is a fit of Eq.~(\ref{eq:leadspectrum}) to the simulated spectrum in the energy range of the reference region 
from 40 to \unit[90]{keV} (solid line), extrapolated to the acceptance region (dashed line).
}
\label{fig:SRIM_spectra}
\end{figure}

Figure~\ref{fig:SRIM_spectra} compares the published SRIM spectrum (digitized from \cite[Fig.~10]{Angloher2009_run32}) 
and our results.
A similar fit function (Eq.~(\ref{eq:leadspectrum})) is used to describe the spectrum in the 40~--~90~keV range, with the resulting 
$E_\text{decay}^\text{Pb}=15.1\pm\unit[1.3]{keV}$ for silver (or $11.8\pm\unit[1.0]{keV}$ for bronze).  
There is a good agreement between simulated curves and the decay constants are consistent with the result from the CRESST paper, 
\unit[13.6]{keV}, if the uncertainty of about \unit[1]{keV} on the CRESST constant is taken into account.

\subsection{Contribution from ion sputtering}
A lot of insight can be gained by considering individual contributions to the spectrum in Fig.~\ref{fig:SRIM_spectra}. 
Apart from the obvious contribution from the polonium decay products ($^{206}$Pb ion and $\alpha$-particle), there is also a 
significant fraction of energy carried into the detector by sputtering products.

In the sputtering process atoms are ejected from a base material due to bombardment of the target by energetic particles.
Sputtering is driven by momentum exchange between the ions and atoms in the material.
The incident ions set off collision cascades in the base material. When such cascades recoil and reach the surface with an 
energy exceeding the surface binding energy (typically of the order of a few eV), an atom can be ejected.

An approximated analytical model for describing sputtering in the cascade regime for amorphous flat base materials has 
been developed by Thompson~\cite{Thompson} and a more sophisticated algorithm is included in SRIM. 
The maximum recoil energy available to sputtered atoms is given as $\Lambda E_1$, where $\Lambda = 4 M_1 M_2/(M_1+M_2)^2$,
$M_1$ and $M_2$ are the masses of the incoming and sputtered ions, and
$E_1$ is the energy of the incoming ion. For silver this gives $0.9 E_1$, i.e. up to \unit[93]{keV}, 
for copper, the dominant component of bronze, up to to 75~keV.

Depending on the kinematics, the distance of the collision point from the surface, and the stopping power, 
a significant fraction of the base material atoms can be kicked off the material with a total kinetic energy, typically
distributed among several ions, exceeding \unit[10]{keV}.
The model predicts $1/E^2$ energy dependence for the spectrum of sputtered atoms, in the region between the surface binding energy, $E_b$
(typically of the order of a few eV) and the maximum recoil energy $\Lambda E_1$.

\begin{figure}
 \centering
 \includegraphics[width=\linewidth]{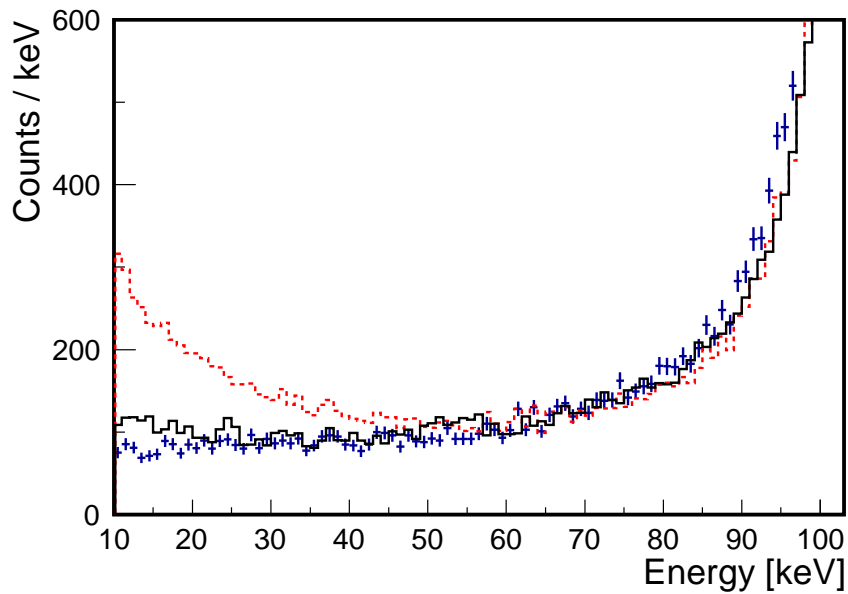}
 \caption{Main components to the background induced by the clamps: 
(black solid line) contributions from $^{206}$Pb recoils alone, (red dashed line) contributions from both $^{206}$Pb and sputtered ions, summed up for each decay, 
and (blue points) the full spectrum, i.e. including also the contribution from $\alpha$-particles (equivalent to Fig.~\ref{fig:SRIM_spectra}).
}
\label{fig:SRIM_sputtering}
\end{figure}

Figure~\ref{fig:SRIM_sputtering} shows the energy spectrum broken down into its main components.
It is evident that taking into account the energy of sputtered atoms introduces a slope increasing towards low energies. 
A significant energy share carried by sputtering products at the low-energy end of the spectrum is an effect that the CRESST collaboration 
has investigated both experimentally and by means of SRIM simulations~\cite{Westphal2008, WestphalPhD}.

Interestingly, the enhancement at low energies is not seen in the simulated full spectrum, as the low-energy events dominated by 
the ion sputtering products are typically accompanied by a high-energy $\alpha$-particle entering the detector, as will be discussed later. 
This effectively moves the events out of the region of interest towards higher energies (and also towards higher light yields, as the recoil scintillation is significantly more quenched than the alpha scintillation).

\subsection{Surface roughness}
As concluded in the previous section, the flat shape of the spectrum of $^{210}$Po decay products at low energies is an outcome of a specific balance 
between sputtered atoms and $\alpha$'s entering the detector. So far, it has been assumed that the clamp surface is perfectly smooth. 
Since ranges of both types of particles in silver vary dramatically ($\sim$~10~microns for $\alpha$'s 
vs. \unit[16]{nm} for recoils) it is easy to imagine that taking realistic surface roughness into account will affect them differently and break that balance,
with some consequences for the acceptance region.
 
SRIM does not allow for treatment of arbitrary geometries and rough surfaces, contrary to \geant~\cite{Geant4}, a package for simulation 
of the passage of particles through matter, developed by the high-energy physics community. \geant\ is a flexible framework, with increasing
usefulness also for low-energy physics applications. In particular, one of its extensions, available as one of its extended examples 
(extended/electromagnetic/TestEm7)
from within the standard distribution, contains all physics relevant for multiple interatomic and alpha scattering in \unit[10]{keV} -- \unit[10]{MeV} energy 
range~\cite{Screen, G4Screen}. The code is actually meant ``to bridge the gap between the effective handling of low-energy processes in simple geometries provided by
SRIM and the very general framework for nuclear events and complex geometries already available in \geant'' (see Ref.~\cite{G4Screen}). It has been extensively benchmarked against
SRIM with respect to nuclear straggling and implantation problems, as well as backscattering, with satisfactory agreement.

A simplified geometry of the problematic clamp region was implemented as a bronze or silver volume with a $10\times10$ micron rectangular area and a 
few micrometers thick, located just next to a ``detector'' volume of a larger area. 
Decays of $^{210}$Po are generated in the material (of the same composition as in SRIM) according to the same exponential depth distribution,
and then all decay products and secondary cascades are tracked and registered when at the boundary of the detector volume.

\begin{figure}
 \centering
 \includegraphics[width=\linewidth]{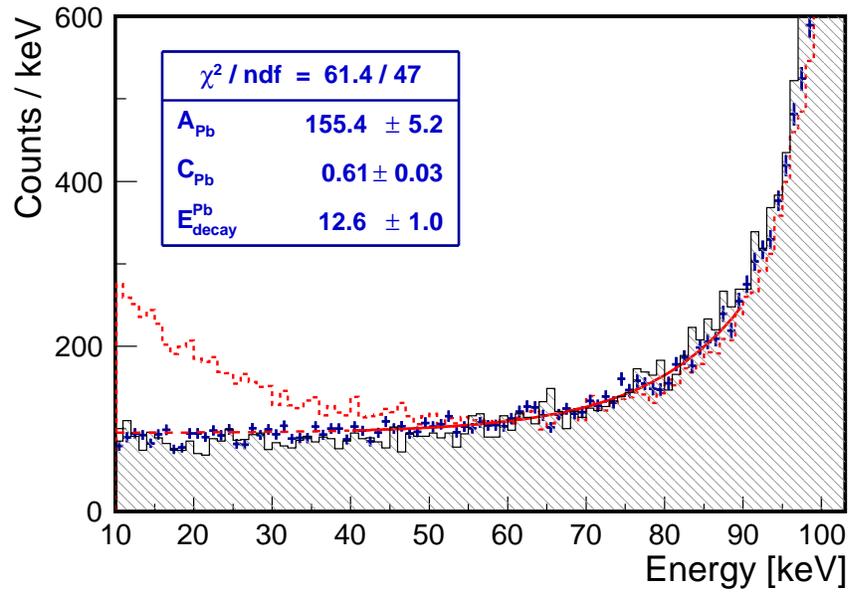}
 \caption{\geant\ simulation results for a perfectly smooth silver surface: (blue points) full energy spectrum from all contributions, 
(red dashed histogram) contributions only from both $^{206}$Pb and sputtered ions, summed up for each decay, 
compared to SRIM simulation result digitized from Fig.~10 in the CRESST paper (shaded black histogram). 
The red curve is a fit of Eq.~(\ref{eq:leadspectrum}) to the simulated blue spectrum in the
energy range of the reference region from 40 to \unit[90]{keV} (solid line), extrapolated to the acceptance region (dashed line).}
\label{fig:geant4_bronze}
\end{figure}

Initially the simulation was repeated for a flat clamp surface, with results in good agreement to SRIM, shown in Fig.~\ref{fig:geant4_bronze}.

\begin{figure}
 \centering
 \subfloat[]{\label{fig:alpharec}\includegraphics[width=0.55\linewidth]{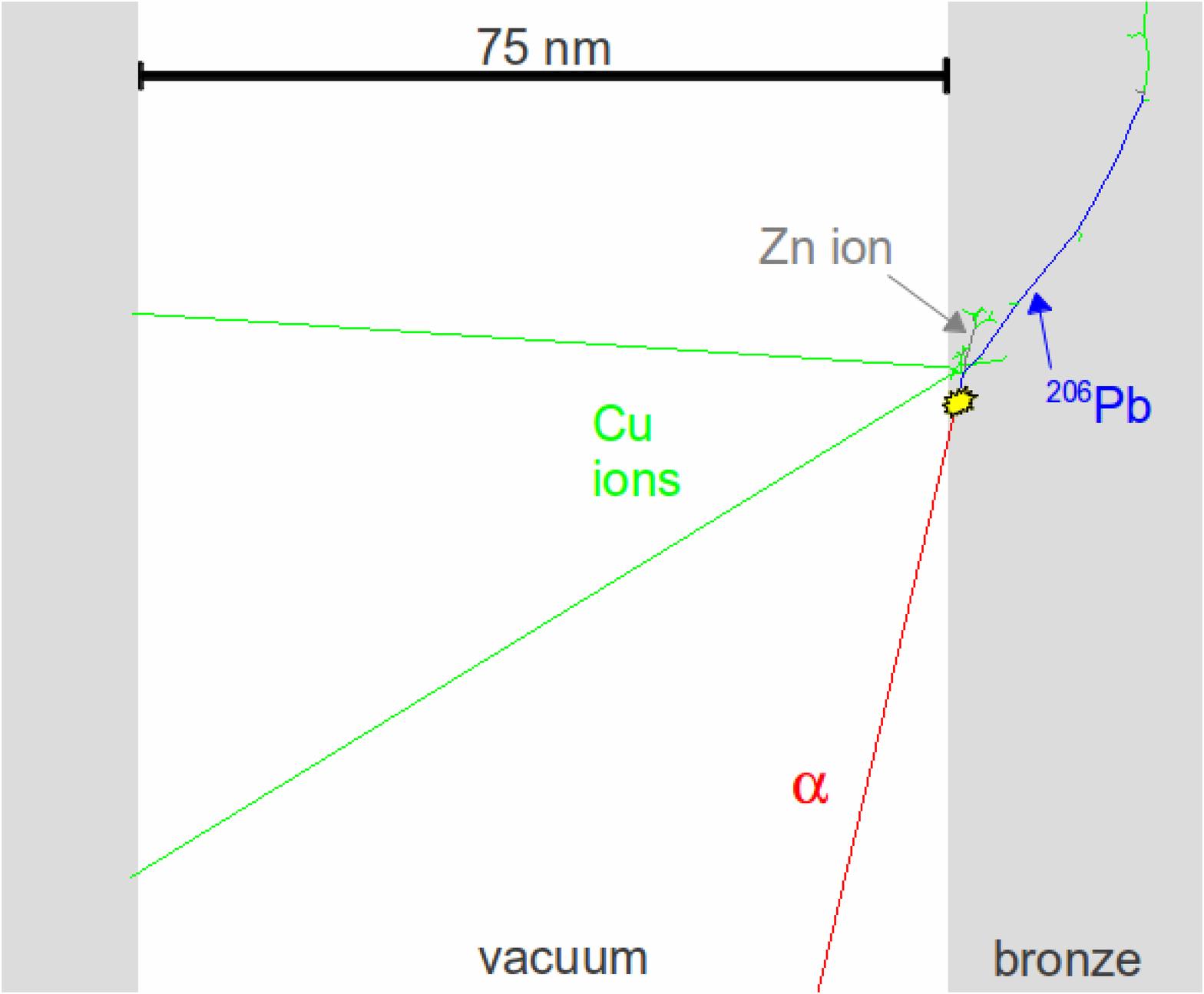}}~
 \subfloat[]{\label{fig:noalpharec}\includegraphics[width=0.39\linewidth]{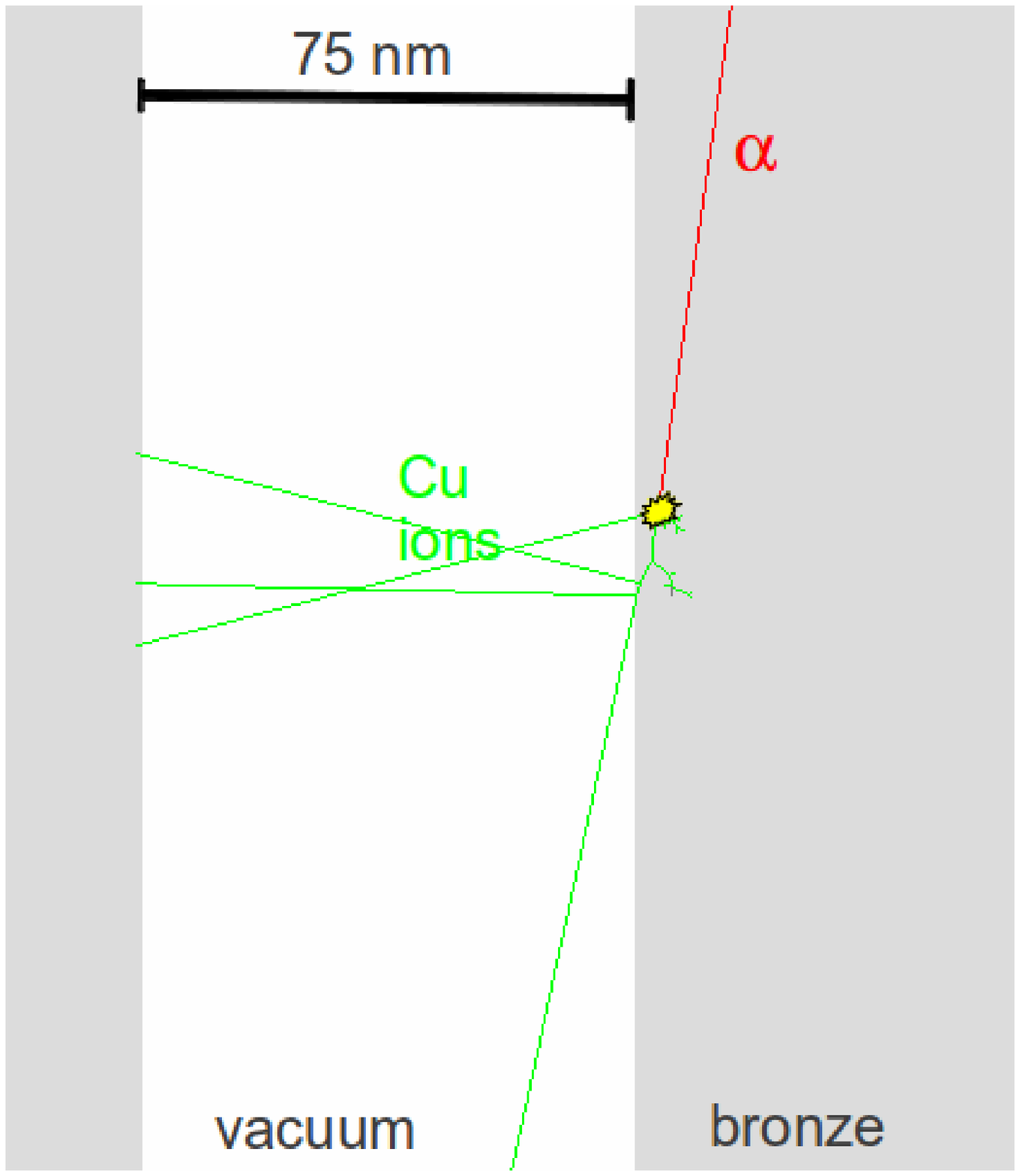}}

 \subfloat[]{\label{fig:geant4_geom}\includegraphics[width=0.4\linewidth]{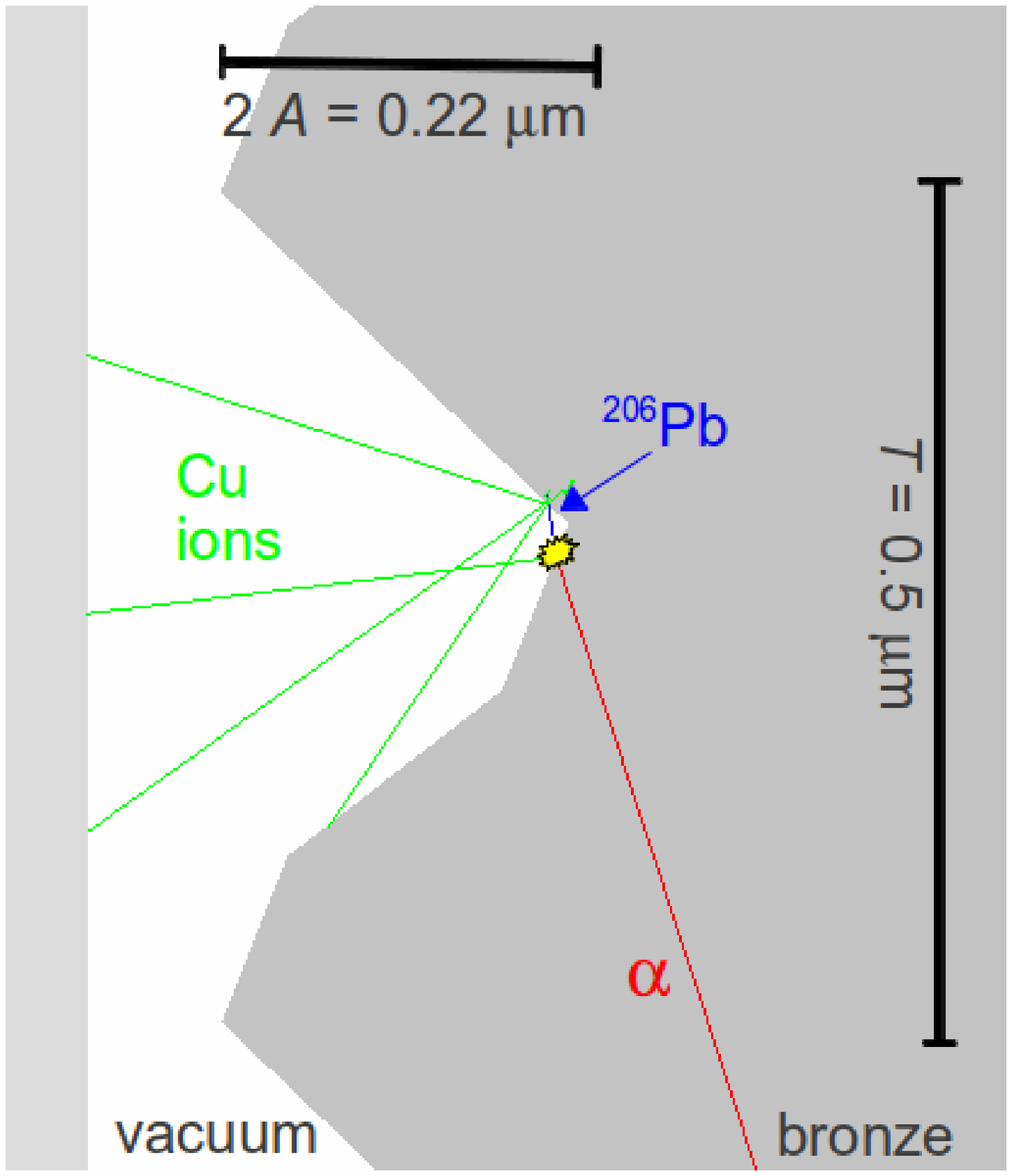}}
 \caption{Zoomed-in cross section of \geant\ geometries showing relevant event types: (a) a typical sputtering-dominated event on a flat surface; 
Cu recoils carrying 10~--~40~keV of energy, with a contribution from a high-energy $\alpha$ (which will eventually reach the detector and shift the event out
of the acceptance region), 
(b) a cascade of sputtered Cu ions reaching the detector (total energy above 40~keV), resulting in an event with no contribution from $\alpha$
and (c) an example of a rough surface (case~\#7), with a part of the 
cascade stopped by surface irregularities, which effectively shifts the event towards lower energies.
$\alpha$'s are shown in red, $^{206}$Pb in blue and copper recoil nuclei in green. Electrons are not shown. Particles are ``detected'' and killed after reaching the surface on the left side of the plot (which extends much beyond the plot boundaries).}
\end{figure}

As for the actual surface roughness, a simplified model was used, which assumed perfect sine-like waviness in one direction. Directionality of surface roughness is, 
in fact, a frequent outcome of machining or polishing.
The surface geometry was implemented using a standard \geant\ volume (G4Polyhedra), which forced us to approximate the sine-like waviness with finite length
linear sections, as shown in Fig.~\ref{fig:geant4_geom}.

Generally, a completely arbitrary surface profile could be implemented in a similar way. It would also be possible to use a 2D surface profile, directly measured
e.g. with a profiler or an atomic force microscope (AFM).

Simulations were run for some arbitrarily chosen values parametrizing the roughness, shown as cases~\#1~--~\#4 in Table~\ref{tab:roughness}, 
where ``waviness'', $T$, is simply the period of the sine function describing the surface and the average roughness, $R_a$, is defined as: 
\[ R_a = \frac{1}{T} \int_{0}^{T}\left|A\cdot\sin\left(\frac{2\pi}{T}x\right)\right|\dd x = \frac{2 A}{\pi}.\]

More realistic sets of values are studied in cases~\#5~--~\#8. Based on literature data found on the surface roughness 
of an ``apparently smooth'' bronze surface~\cite{Quiney} and the roughness of electroplated 
silver coatings~\cite{Foster2003,Foster2005}, we ran simulations for the average roughness of 0.07~micrometers (2.7~$\mu$inch),
which was reported for bronze and is also consistent with the smoothest silver coating reported in Ref.~\cite[Fig.~5]{Foster2003}. 
The relevant range of the waviness parameter (0.1--5 micrometers) was specified based on AFM surface scans from \cite{Foster2003, Foster2005}. 
 
\begin{figure}
 \centering
 \begin{overpic}[width=0.5\linewidth]{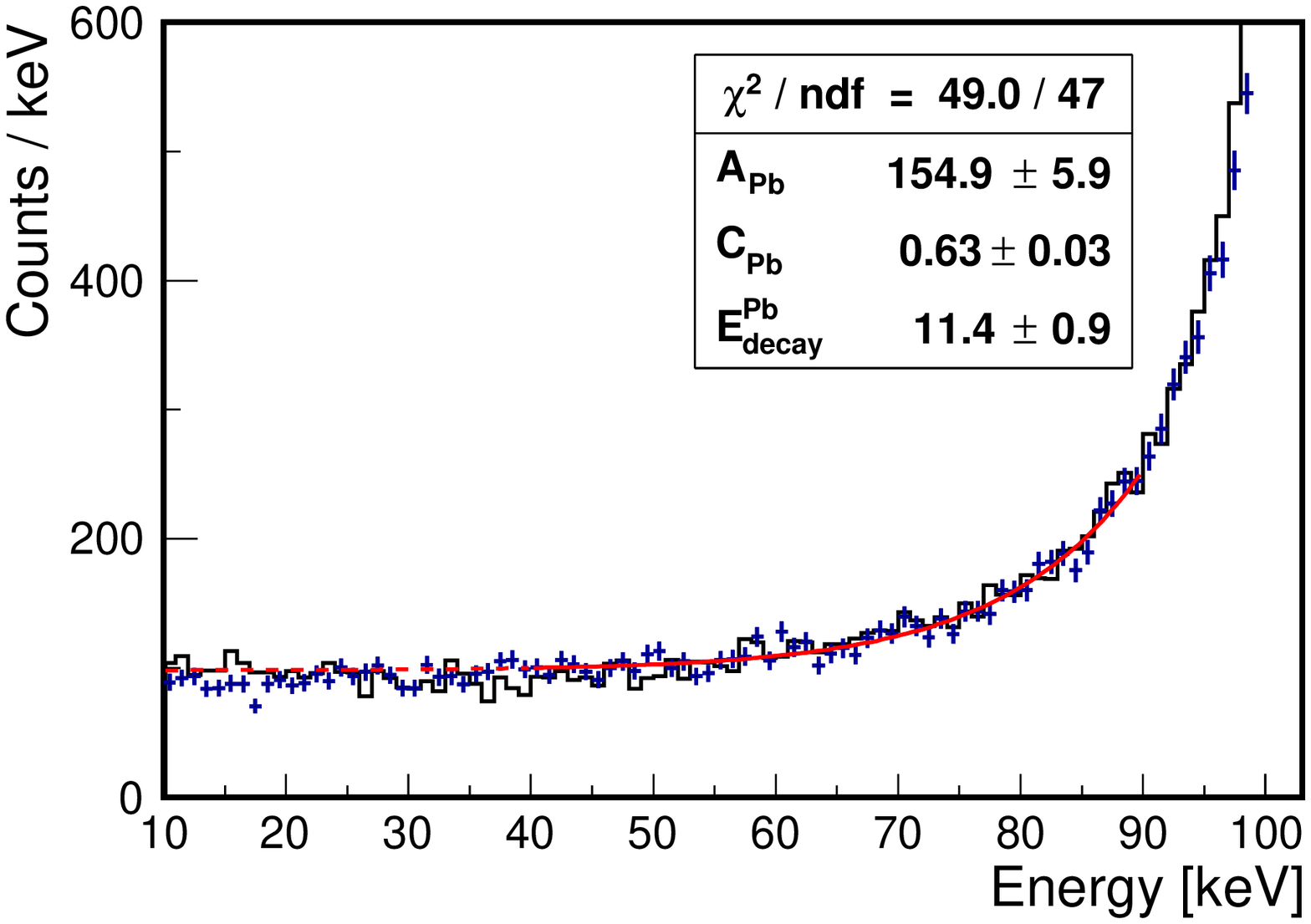}\put(21,55){\tiny Silver \#5}\end{overpic}\begin{overpic}[width=0.5\linewidth]{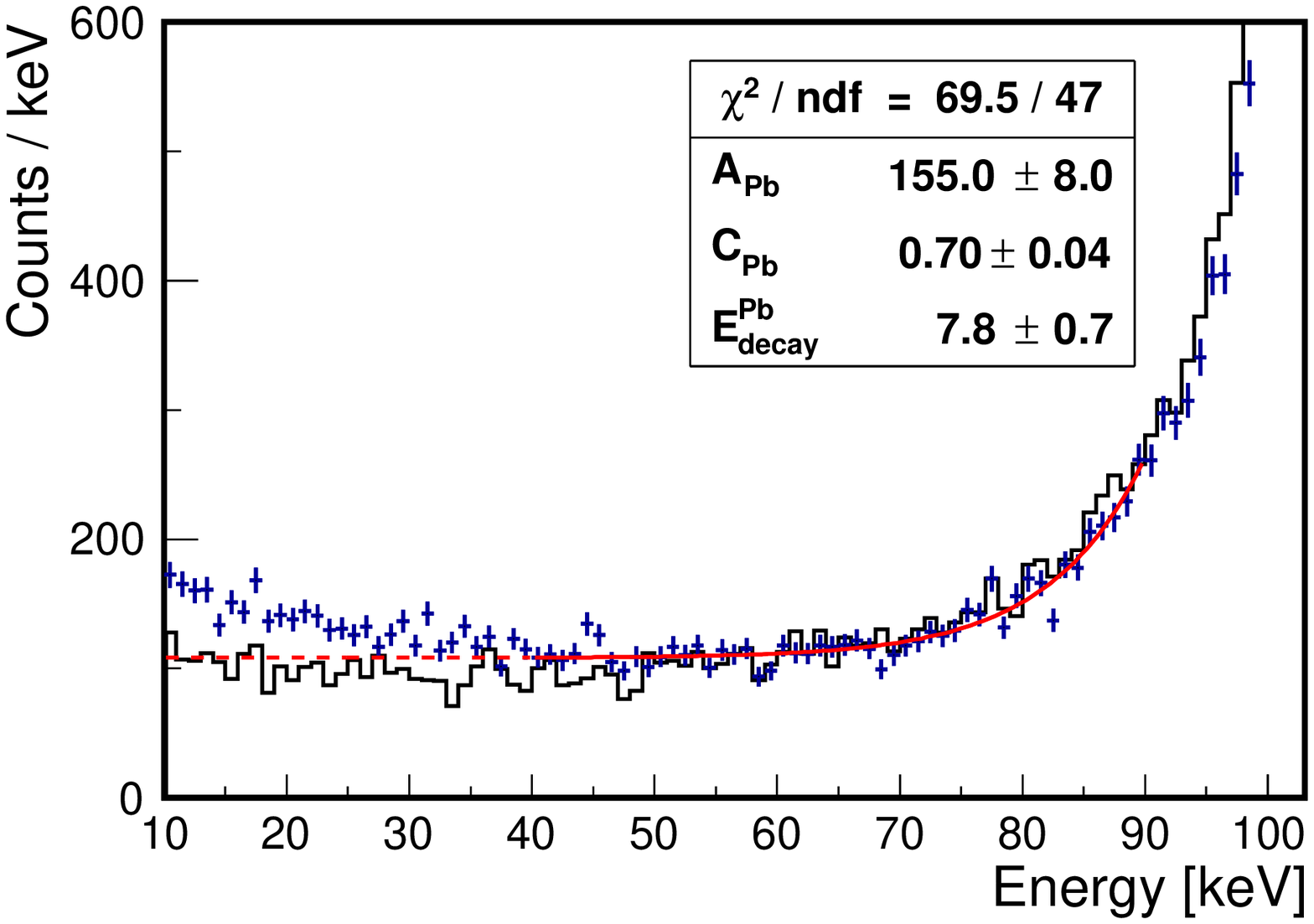}\put(21,55){\tiny Silver \#6}\end{overpic}

 \begin{overpic}[width=0.5\linewidth]{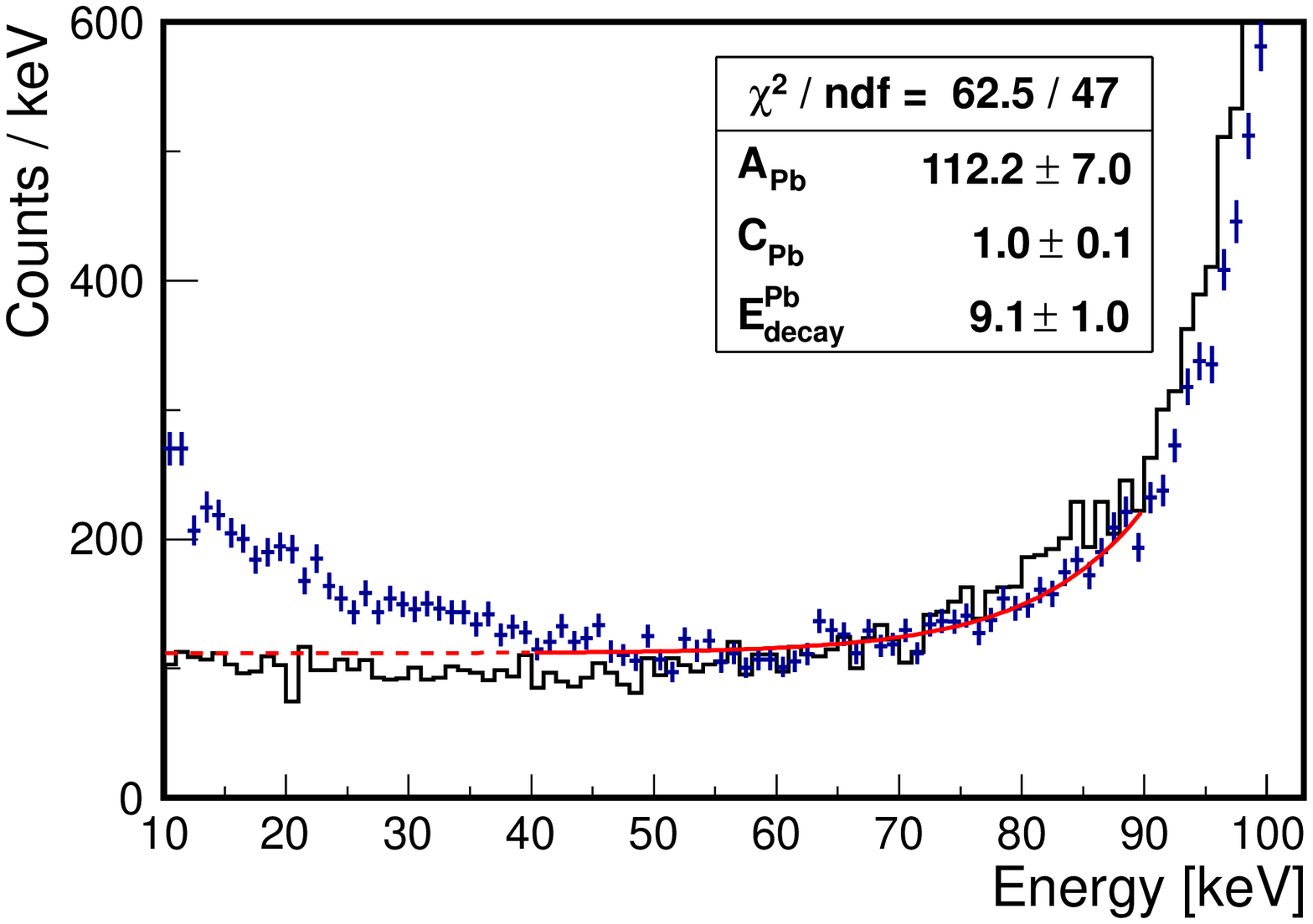}\put(21,55){\tiny Silver \#7}\end{overpic}\begin{overpic}[width=0.5\linewidth]{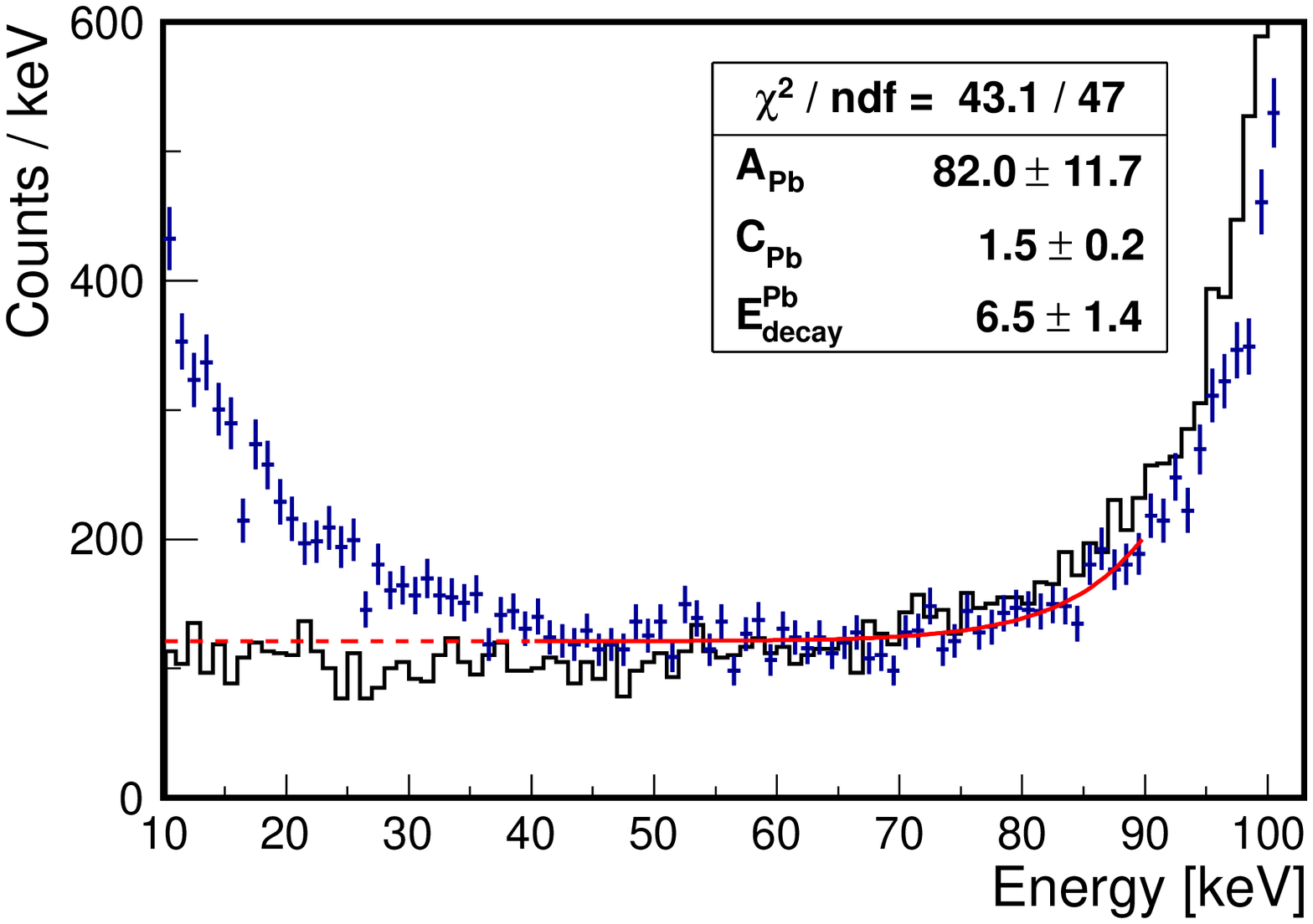}\put(21,55){\tiny Silver \#8}\end{overpic}
 \caption{\geant\ simulated spectra for all respective ``rough'' cases from~Table~\ref{tab:roughness}. Plots contain:
(blue points) full energy spectrum from all contributions, 
(black solid line) energy from $^{206}$Pb recoils only (no sputtering and no alphas included).
The red curve is a fit of Eq.~(\ref{eq:leadspectrum}) to the simulated blue spectrum in the
energy range of the reference region from 40 to \unit[90]{keV} (solid line), extrapolated to the acceptance region (dashed line).}
\label{fig:geant4_10A5T}
\end{figure}
As mentioned earlier, there is no statistically significant difference in results between bronze and silver. Qualitatively, 
all but one simulated configuration show the same effect, namely, the excess of events at lower energies, when compared to the ``flat'' surface model.
Resulting spectra for all rough cases from~Table~\ref{tab:roughness} are shown in Fig.~\ref{fig:geant4_10A5T}. 

The only exception from the rule is case~\#5, where the low energy contribution disappears, making the result similar to that of a perfectly smooth surface, 
i.e. case~\#0, discussed earlier. Some non-trivial dependence on the $A/T$ ratio is not a surprise, since smoother surfaces will have smaller $A/T$ ratios.
This gives an idea on what kind of a surface finish would be required to mitigate backgrounds of this type.
\begin{table*}
\begin{center}
{\footnotesize
  \begin{tabular}{l|c|c|c|c|c|c}
  & \multicolumn{4}{c|}{Surface Roughness} & \multicolumn{2}{c}{Silver}\\
  \hline
     \# & $A$ [$\mu m$]& $T$ [$\mu m$] & $A/T$ & $R_a$ [$\mu m$] & $E_\text{decay}^\text{Pb}$ [keV] & $n_\text{ref}^\text{Pb} / n_\text{acc}^\text{Pb}$\\ \hline
     0 & 0 & -- & -- & 0 & $12.6\pm1.0$ & 2.46 \\ \hline
     1 & 1 & 5.0 & 0.2 & 0.64 & $7.8\pm0.9$ & 1.29 \\
     2 & 1 & 2.5 & 0.4 & 0.64 & $6.2\pm0.9$ & 1.02 \\ 
     3 & 10 & 5.0 & 2 & 6.4 & $4.7\pm1.3$ & 0.89 \\
     4 & 0.020 & 0.005 & 4 & 0.013 & $9.5\pm1.4$ & 1.29 \\ \hline
     5 & 0.11 & 5.0 & 0.022 & 0.07 & $11.4\pm0.9$ & 2.45 \\
     6 & 0.11 & 1.0 & 0.11 & 0.07 & $7.8\pm0.7$ & 1.67 \\
     7 & 0.11 & 0.5 & 0.22 & 0.07 & $9.1\pm1.0$ & 1.31 \\
     8 & 0.11 & 0.1 & 1.1 & 0.07 & $6.5\pm1.4$ & 1.1 \\
  \end{tabular}
}
\end{center}
\caption{Roughness for all cases used in \geant\ simulations: amplitude ($A$), waviness ($T$), their ratio, and average surface roughness ($R_a$). 
Two last columns summarize the simulation results and contain, respectively: exponential decay constant fitted to the reference region ($E_\text{decay}^\text{Pb}$), 
and the ratio of number of counts in the reference region to the number of counts in the acceptance region ($n_\text{ref}^\text{Pb} / n_\text{acc}^\text{Pb}$) for silver. Statistically consistent results obtained for bronze are omitted.
Case \#0 corresponds to data shown in Fig.~\ref{fig:geant4_bronze}.
}
\label{tab:roughness}
\end{table*}

We did not see any similar effect of difference between a smooth and a rough surface for $\alpha$'s generated deeper in the clamps, 
i.e. uniformly from a 15~micrometer thick surface layer. This can be explained by the range of $\alpha$'s in bronze or silver, 
which is much larger than the range of $^{206}$Pb recoils and also larger than the typical roughness scale. 

\subsection{Light-quenching factors}
So far we have not considered the consequences of light-quenching factor (QF) difference between $^{206}$Pb and
the sputtered ions. The QF is defined as a light output of a particle, relative to the light output of a $\gamma$
of the same deposited energy. Nuclear recoil quenching factors in CaWO$_4$ decrease with increasing atomic mass
of an ion~\cite{QF,QF2}. CRESST~\cite{Angloher2009_run32} reported values for Ca and W of 6.4\% and 3.9\%,
respectively, and 1.4\% for Pb (all values with about 10\% relative uncertainty). Based on~\cite[Fig. 1]{QF2} one can infer
the QF value for silver of around 3\%.

Therefore, the low energy part of the spectrum, dominated by the sputtered silver ions, should have about 2~--~3 times
larger light yield than Pb ions. This is roughly consistent with Figure 12 in the CRESST paper~\cite{Angloher2009_run32},
where the light yield distribution of a possible WIMP signal, corresponding to the likelihood maximum M1,
is centered around 5\%.

\section{Full model and the CRESST-II results}
Below, in Fig.~\ref{fig:SRIM_data}, we present a direct comparison of the low energy spectrum measured by CRESST with the spectra obtained from \geant\ simulations
taking the surface roughness effects into account. 
\begin{figure}
 \centering
 \begin{overpic}[width=1\linewidth]{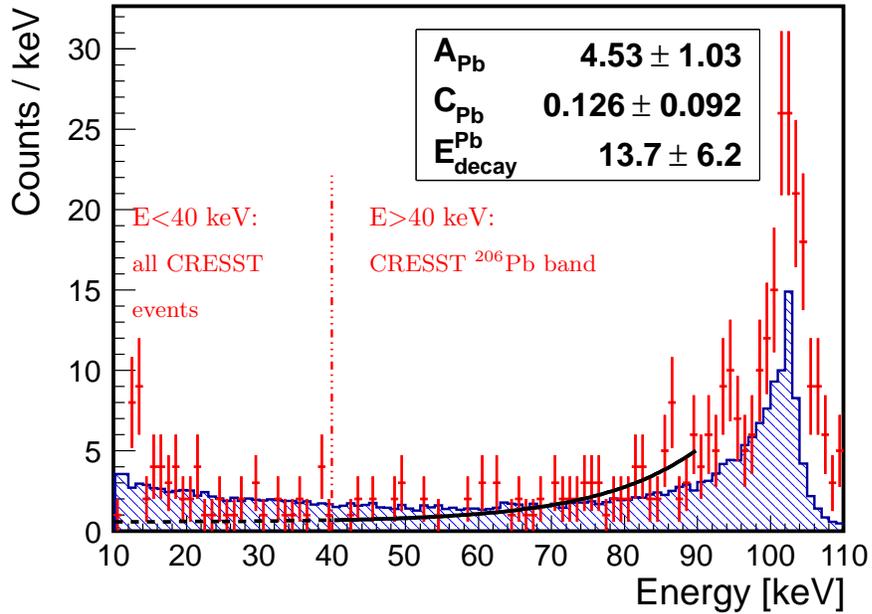}
   \put(45,45){\textcolor{red}{E$>$40 keV:}}\put(19,45){\textcolor{red}{E$<$40 keV:}}
   \put(45,40){\textcolor{red}{\small CRESST $^{206}$Pb band}}\put(19,40){\textcolor{red}{\small all CRESST}}\put(19,35){\textcolor{red}{\small events}}
 \end{overpic}
 \caption{Comparison of CRESST data (red points) with CRESST's clamp background model (black curve) and our model including surface roughness 
(blue shaded histogram). The CRESST data was digitized from Figures 9 and 11 in Ref.~\cite{Angloher2009_run32}. Our \geant\ simulation of the 
clamp background spectrum corresponds to case~\#7 (see text) and was normalized to give the same integrated number of 89 counts in the reference 
region as the CRESST data. The CRESST background model is based on a log-likelihood fit of Eq.~(\ref{eq:leadspectrum}) to the data in the
reference region (40 to \unit[90]{keV}, shown as solid line). Extrapolation of the fitted function to the acceptance region (10~--~\unit[40]{keV}, shown as 
dashed line), is consistent with what was used in the CRESST analysis to estimate the clamp background contribution.}
\label{fig:SRIM_data}
\end{figure}

The exponential fit to the closest matching spectrum, which corresponds to case~\#7, yields $9.1\pm\unit[1.0]{keV}$ for the exponential decay constant.
As the uncertainty on the slope parameter of the exponential fit to the reference region has not been included in the CRESST paper, we repeated the fit, 
confirming the published values, in particular, $E_\text{decay}^\text{Pb}=13.72\pm\unit[6.19]{keV}$\/. It is, thus, within error consistent with
the slope of the ``full cascade'' spectrum.

We note, that the full energy $^{206}$Pb recoil peak at \unit[103]{keV} from the simulation has lower intensity, 
when compared with the CRESST spectrum. Out of many possible explanations, perhaps the most likely ones are the surface component of $^{210}$Po contamination exceeding the assumed exponential depth distribution or surface contamination of the CaWO$_4$ crystal directly under the clamps.

However, most importantly, we find a fair agreement between the integrated number of counts in the acceptance region between our model 
(68 counts or 53 counts, based on case~\#7 or case~\#6, respectively) 
and the CRESST data, i.e. 67 counts from all backgrounds (which includes very approximately 40~--~45 counts 
attributed in the global likelihood fit to the WIMP signal and the $^{206}$Pb recoils taken together).

Based on our qualitative study, the clamp background is significantly larger than estimated in the CRESST background model, and might be the 
dominant contribution to the low-energy part of the spectrum. It remains unclear, however, whether it is sufficient to entirely explain the excess 
of events, which necessitates further, better informed, analysis.

In addition to this, we note that it is likely that the same class of low-energy events correlated with $^{206}$Pb contamination of the clamps, 
contributed significantly to the earlier spectra published by CRESST as Figures~3 and 8 in Ref.~\cite{CRESSTrecoil}, which exhibit similar event 
excesses at lower energies (attributed by the authors entirely to neutrons, for the former, and $e/\gamma$-band leakage, for the latter).

\section{Summary}
We have found a significant contribution to the number of events in the acceptance region in CRESST-II data coming from cascades of secondary nuclear recoils 
caused by recoiling \unit[103]{keV} $^{206}$Pb nucleus from decays of $^{210}$Po present on and under the surface of clamps, 
used to hold the target. The size of this contribution becomes significant only when a realistic model of surface roughness is included in the calculation.

This is a new type of background which to our knowledge has not been previously considered in the context of ultra-low background experiments. 
It is, however, fairly easy to model using some commonly used simulation tools, such as \geant. The study itself has been motivated by 
similar surface roughness effects discovered in data from DEAP-1~\cite{DEAP-1}, a prototype for the DEAP-3600~\cite{DEAPpaper} dark matter detector 
with a single-phase liquid argon target.

Based on the published data, we have qualitatively estimated the expected number of this class of events in the CRESST-II
dataset, and found it approximately consistent with the excess of events in the acceptance region. The expected light yield
of sputtered silver ions is also consistent with the light yield distribution of these events. 
Although a more sophisticated study including a corrected maximum likelihood analysis is necessary to reach the final conclusion, 
and also specific input regarding the clamp composition, nature of contamination and surface roughness would be useful,
we interpret this as an indication that the $^{206}$Pb recoils could be responsible for the dominant part of the signal observed by CRESST-II.

\section*{Acknowledgements}
We thank Prof. Franz Pr\"obst for discussions during his visit at Queen's University, which directly inspired this work. His valuable comments 
and suggestions to an early version of this paper are gratefully acknowledged. We also thank Prof. Wolfgang Rau for fruitful discussions, and 
Svatoslav Florian for suggestions on typical roughness of metal surfaces. This work was supported by the National Sciences and Engineering
Research Council of Canada.

\end{document}